\documentclass[a4paper,11pt]{article}
\usepackage{pos}
\usepackage{subfigure}
\usepackage{slashed}

\title{Higgs decay into two leptons and a photon revisited}
\author*[a]{Aliaksei Kachanovich}
\affiliation[a]{TTP, KIT,\\
Wolfgang-Gaede-Straße 1, Karlsruhe, Germany}

\emailAdd{aliaksei.kachanovich@kit.edu}
\abstract{ I present new results for the Standard-Model predictions of
  the differential decay rates for $H\to \ell^{+} \ell^{-} \gamma$,
  where $\ell=e, \mu$, and the forward-backward asymmetries 
  defined in terms of the flight direction of the
  photon corresponding to the lepton momenta. The results 
  dependend on the cuts on energies and invariant masses of the final
  state particles. For standard choices of these cuts the branching ratios
  $B(H\to e \bar e \gamma)=5.8\cdot 10^{-5}$ and
  $B(H\to \mu \bar \mu \gamma)=6.4\cdot 10^{-5}$ as well as the
  forward-backward asymmetries $\mathcal{A}^{(e)}_{\text{FB}}=0.343$ and
  $\mathcal{A}^{(\mu)}_{\text{FB}}=0.255$ have been found.  }

\FullConference{%
  40th International Conference on High Energy physics - ICHEP2020\\
  July 28 - August 6, 2020\\
  Prague, Czech Republic (virtual meeting)
}

\begin{document}
\maketitle

\section{Motivation}
The Standard Model (SM) is a very successful but incomplete
theory. Theories of new physics typically involve extensions
of the minimal Higgs sector of the SM and precision measurements of
production and decay of the 125-GeV Higgs particle $h$ can eventually
reveal virtual effects of such an extended Higgs sector. Couplings of
$h$ can be modified by the mixing of the SM Higgs field with another
scalar field, as in e.g.\ Higgs portal models of Dark Matter (see
\cite{Krnjaic:2015mbs, Kachanovich:2020yhi,
  Filimonova:2019tuy} for recent studies). Another possible 
impact of new physics ---not necessarily related to extended Higgs
sectors--- is the modification of $h$-couplings through loop effects
involving new particles. Such effects are best detectable in processes
which are suppressed in the SM. This situation occurs if the SM process
itself is loop-dominated or involves a tiny tree-level coupling.  
The process $h \to \ell \Bar{\ell} \gamma$, with $\ell$ representing an
electron $e$ or muon $\mu$, are in this category. 
Discrepancies between different results in the literature
\cite{Chen:2012ju, Passarino:2013nka, Dicus:2013ycd,
  Sun:2013rqa, Han:2017yhy} have triggered the 
new study in Ref.~\cite{Kachanovich:2020xyg} which I present here.

All contributions to the process $h \to \ell \Bar{\ell}$ are
proportional to the Yukawa coupling, so that the leptons in the final
state always have opposite chirality, i.e. if one lepton is right-handed
the second one is always left-handed. By contrast, in the process
$h \to \ell \Bar{\ell} \gamma$ both leptons can be in the same helicity
state. The dominant contribution in the case of the final state
$\ell_{L(R)} \bar{\ell}_{L(R)} \gamma$ comes from the one-loop level and
is not vanishing even for $m_{\ell}=0$. As a result, the rate for the
process $h \to e \bar{e} \gamma$ is much larger than the rate of
$h \to e \bar{e}$, while the analogous rates for muons are of the same
order of magnitude
($\mathcal{B} (h \to \mu \bar{\mu} \gamma) = 6.7 \times 10^{-5}$
vs.  $\mathcal{B}(h \to \mu \bar{\mu} ) = 2 \times 10^{-4}$).

\section{Calculations}
\begin{figure}[tb]
\hrule
	\begin{center}
		\subfigure[t][]{\includegraphics[width=0.22\textwidth]{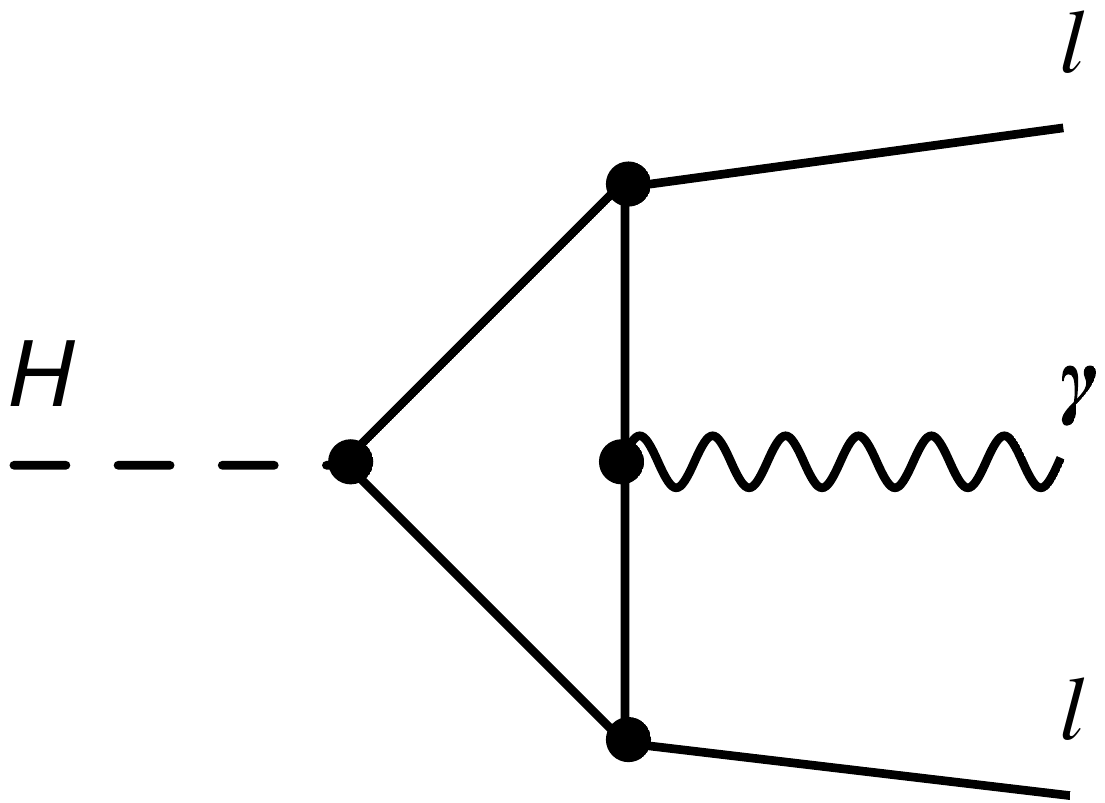}}
		\hspace{.6cm}
		\subfigure[t][]{\includegraphics[width=0.22\textwidth]{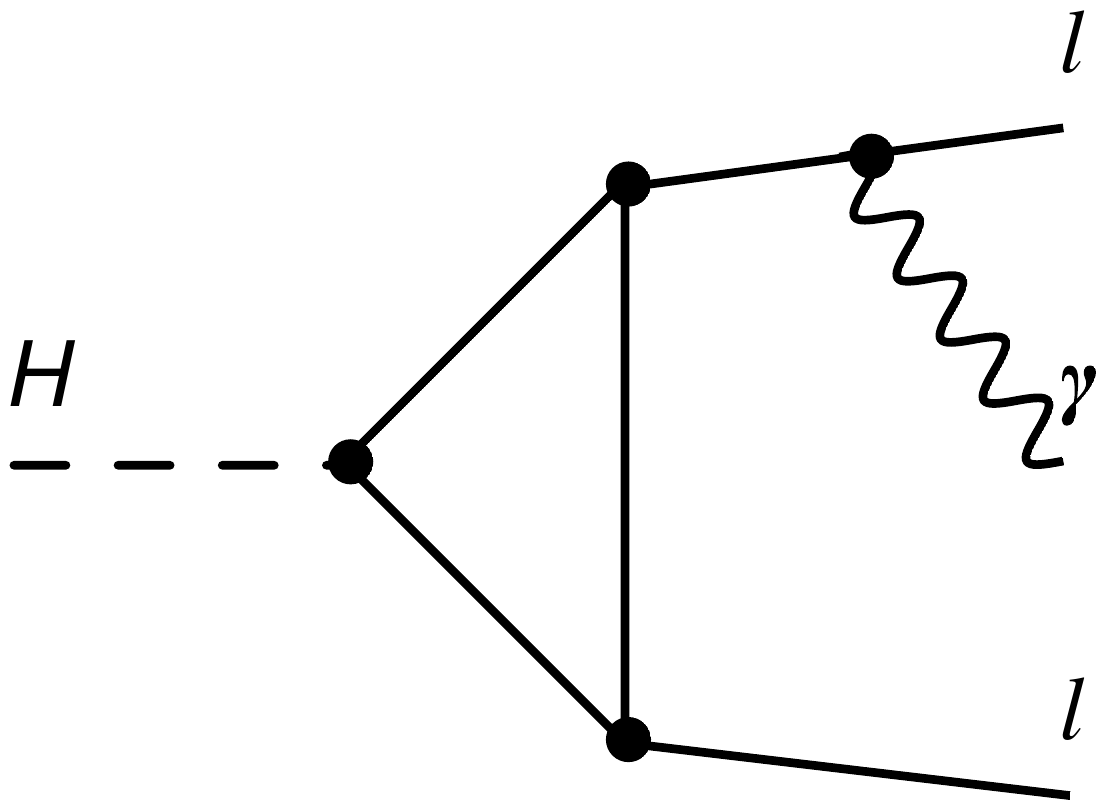}}
		\hspace{.6cm}
		\subfigure[t][]{\includegraphics[width=0.22\textwidth]{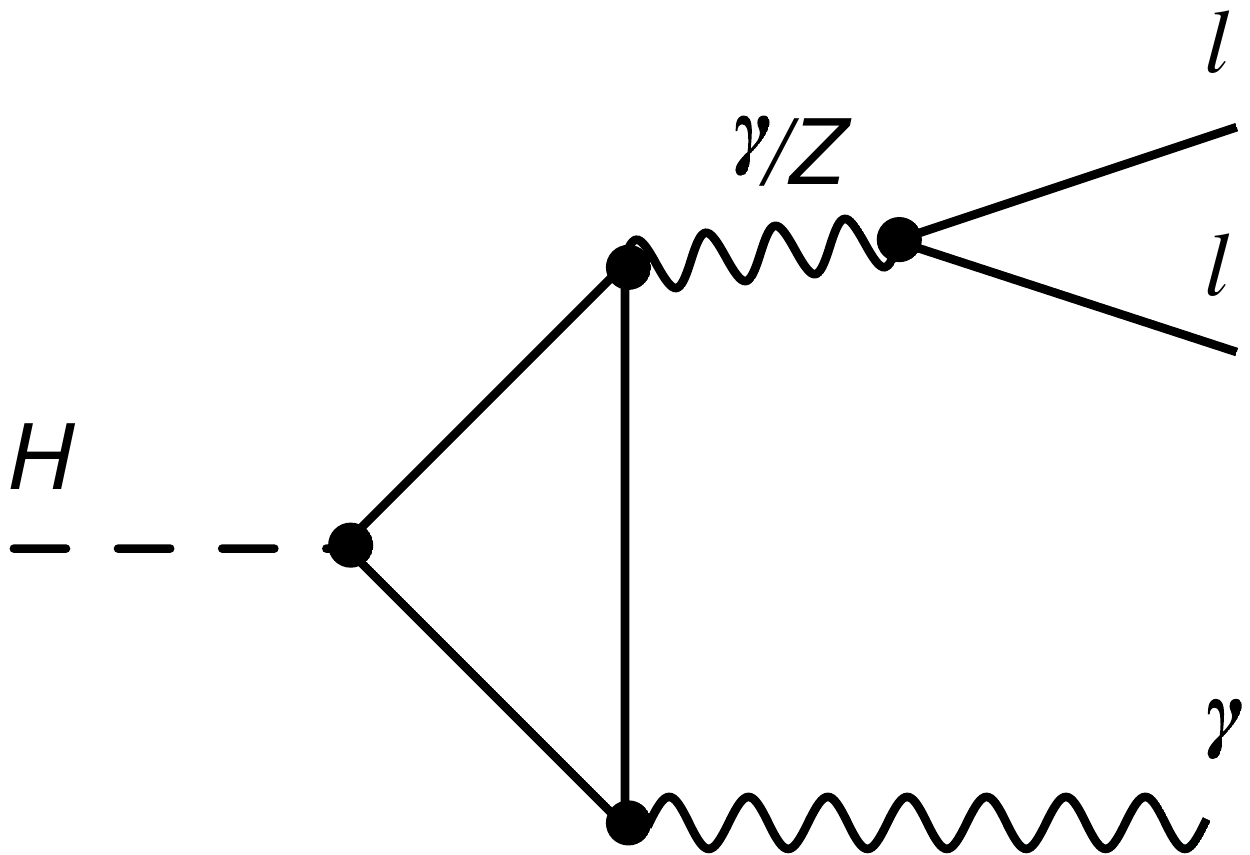}}\\
		\hspace{.6cm}
		\subfigure[t][]{\includegraphics[width=0.22\textwidth]{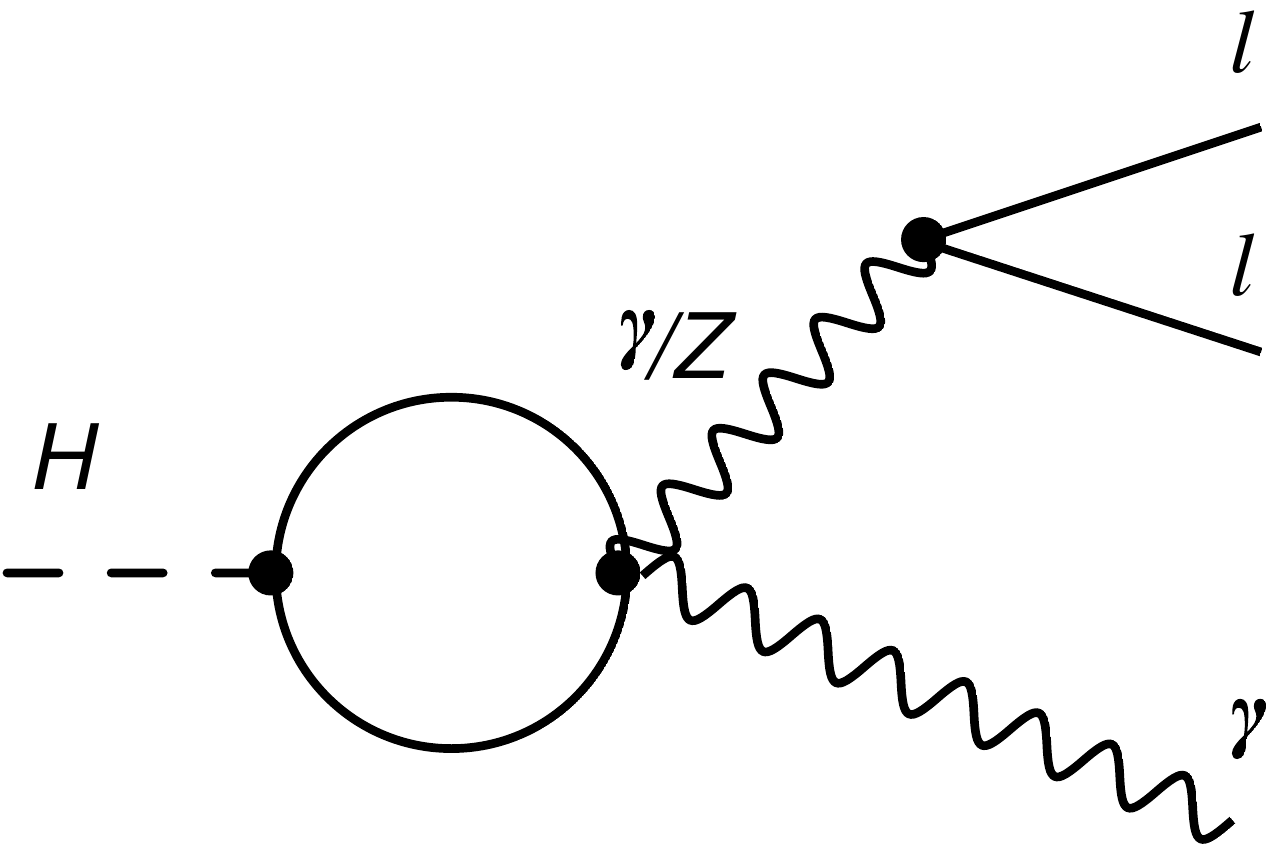}}
		\hspace{.6cm}
		\subfigure[t][]{\includegraphics[width=0.22\textwidth]{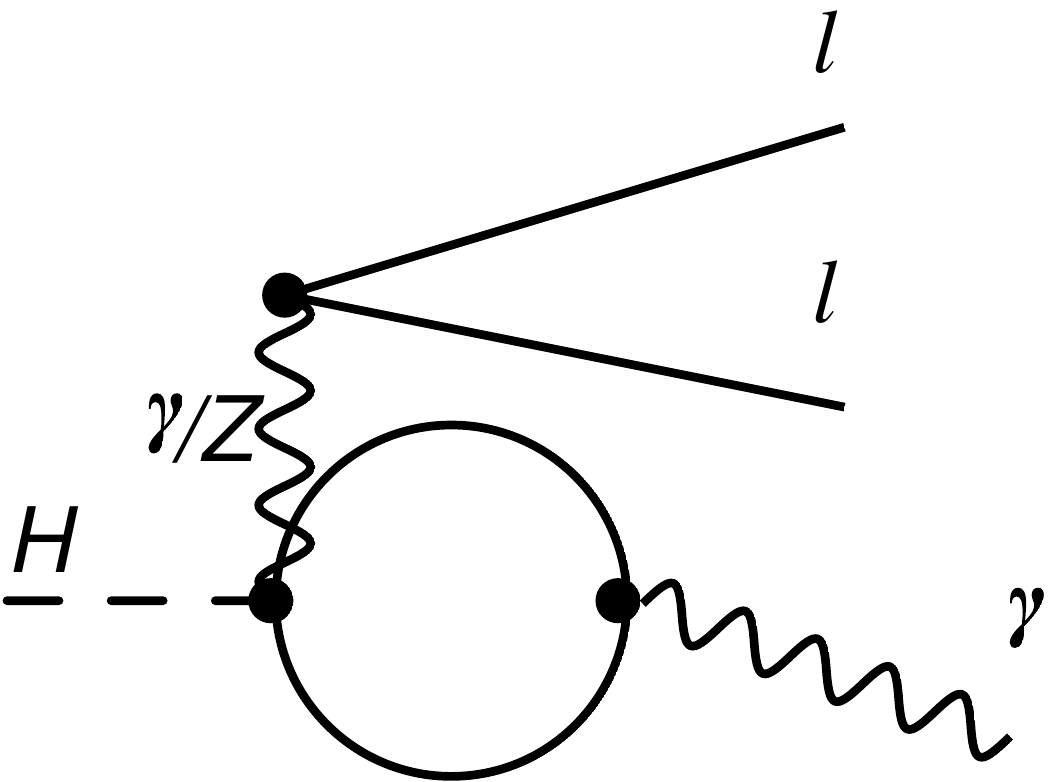}}
		\hspace{.6cm}
		\subfigure[t][]{\includegraphics[width=0.2\textwidth]{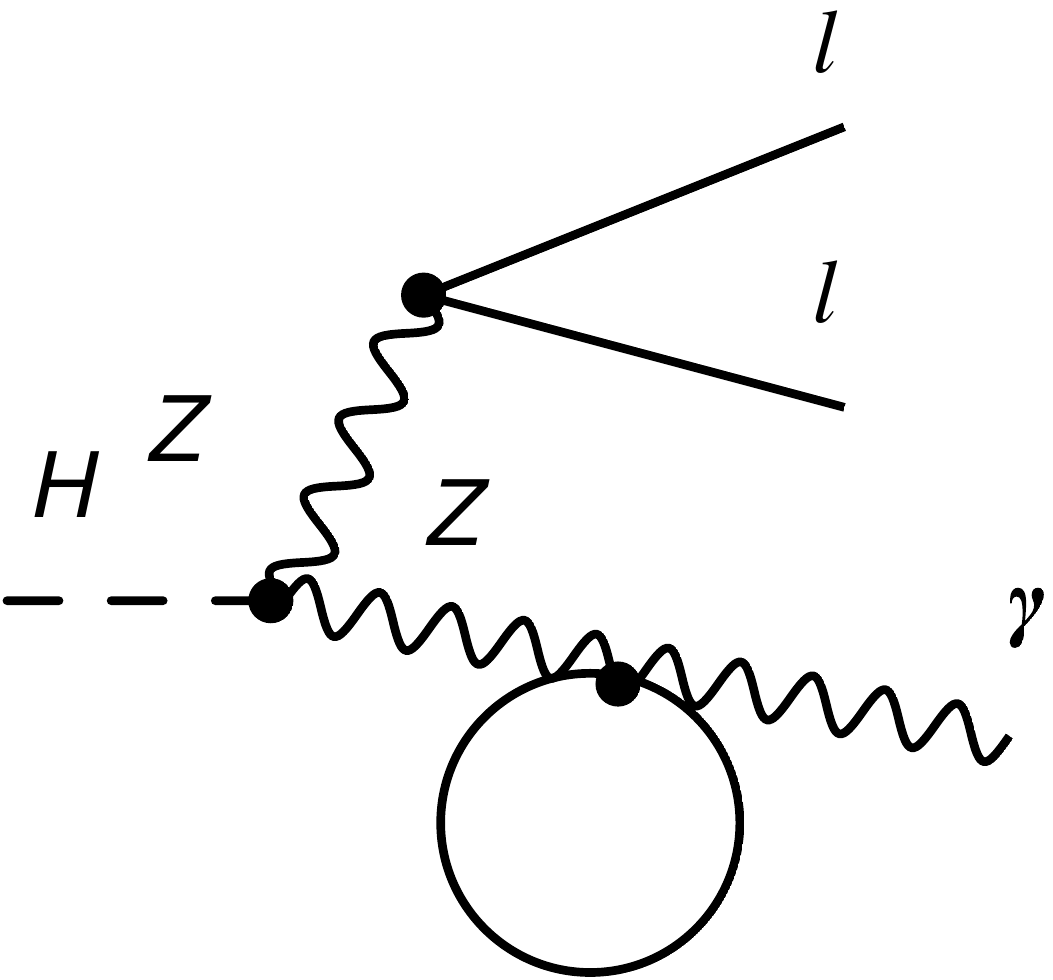}}
		\hspace{.6cm}
		\subfigure[t][]{\includegraphics[width=0.2\textwidth]{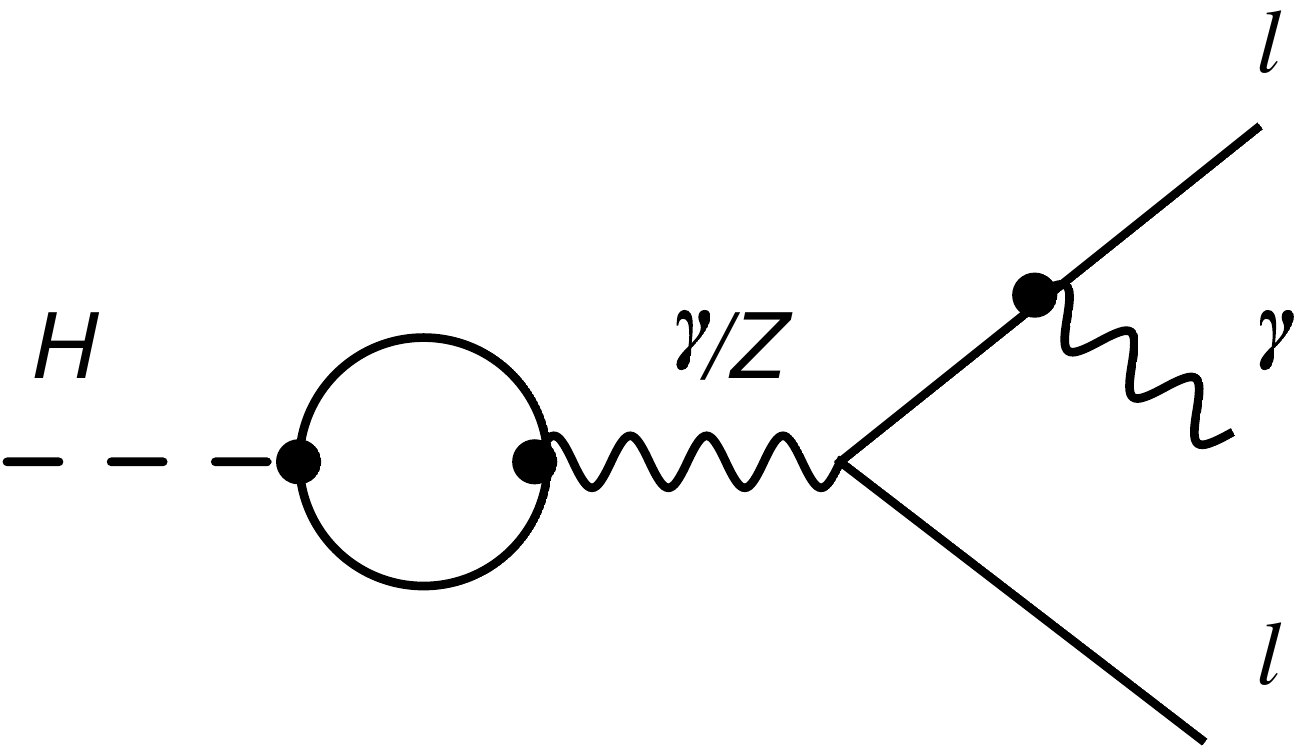}}
		\hspace{.6cm}
		\subfigure[t][]{\includegraphics[width=0.2\textwidth]{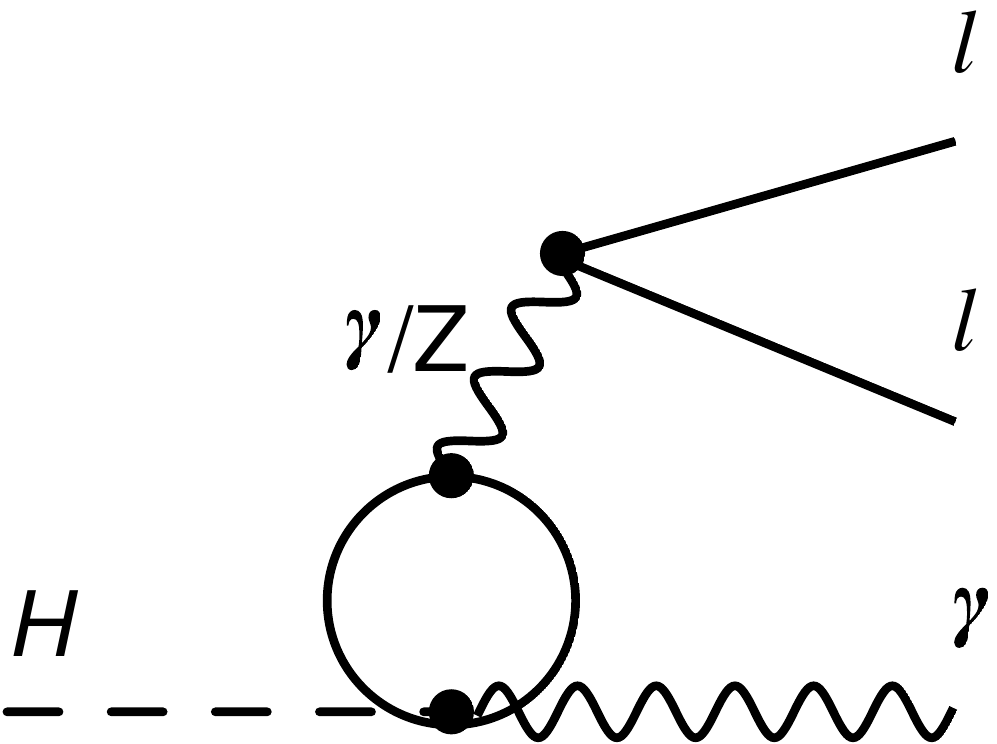}}
         \end{center}
         \caption{These Feynman diagrams represent different classes of
           diagrams which contributes to the process
           $H\to \ell \bar{\ell}\gamma$ at one-loop level. The solid
           lines without label denote generic SM propagators to be
           replaced by all possibilities permitted by the SM Feynman rules.
           Diagrams a) represent Class 1, b) Class 2, c) Classes 3
           and 4, d) Classes 5 and 6, e) Classes 7 and 8, f) Class 9, g)
           Classes 10 and 11, h) Classes 12 and 13. If two classes are
           mentioned, they refer to $\gamma$ and $Z$.  }
	\label{Diagrams}
~\\[-3mm]\hrule
\end{figure}

The \emph{Mathematica} package \emph{FeynArts} \cite{Kublbeck:1990xc,
  Hahn:2000kx} finds 436 one-loop diagrams contributing to
$h \to \ell \Bar{\ell} \gamma$.  Ignoring those involving the small
lepton Yukawa coupling reduces the number of diagrams to 119, which can
be categorised by their topologies and the type of propagator to which
the lepton pair are coupled into 13 classes shown in
Fig.~\ref{Diagrams}. We perform the evaluation of these diagrams with
the help of the \emph{FeynCalc} package \cite{Shtabovenko:2020gxv,
  Shtabovenko:2016sxi, Mertig:1990an} in a general $R_{\xi}$ gauge. For
some analytical checks we used \emph{Package-X} \cite{Patel:2015tea}
which can be linked to FeynCalc via the \emph{FeynHelpers} add-on
\cite{Shtabovenko:2016whf}. One of the goals of this calculation was to
find gauge-independent classes of diagrams. Only Classes 10 and 11 (see
Fig.~\ref{Diagrams}) are gauge independent, while all other diagrams
must be added for another gauge independent subset.  For photon, Z-, and
W-boson propagators three different gauge parameters $\xi_{\gamma}$,
$\xi_{Z}$, and $\xi_{W}$ are used. The cancellation of the parameters
$\xi_{\gamma}$ and $\xi_{Z}$ happens after the summation of all
diagrams. For the cancellation of the parameter $\xi_{W}$, it is crucial
that the W-boson mass $m_{W}$ and the Z-boson mass $m_{Z}$ fulfill the
Weinberg relation
\begin{equation}\label{Eq.:fine structure constant}
    e^{2}/g_{2}^{2}=\sin^{2} \theta_{W}= 1 - m_{W}^{2}/m_{Z}^{2}\,, 
\end{equation}
\noindent where $e$, $g_{2}$ and $\theta_{W}$ are the electromagnetic
and weak coupling constants and the weak mixing angle,
respectively. This condition restricts the choice of the value of the
fine structure constant $\alpha = e^{2}/(4 \pi)$ after $g_2$ is fixed in
terms of $M_W$ by the experimental value of the Fermi constant $G_F$.

The amplitude satisfies the QED Ward identity and can be written as:
\begin{eqnarray}
  \mathcal{A}_{\text{loop}}&=&\big[(k_{\mu}\, p_{1\nu}-g_{\mu\nu}\,
                                k\cdot p_{1})  \bar{u}(p_{2})\big(a_{1}
                                \gamma^{\mu} P_{R} 
   + b_{1} \gamma^{\mu} P_{L}\big) v(p_{1})\nonumber\\
                            &+&(k_{\mu}\,p_{2\nu}-g_{\mu\nu}\, k \cdot p_{2})
                                \bar{u}(p_2)
 \big(a_2\gamma^\mu P_{R} + b_{2}\gamma^{\mu} P_{L} \big)v(p_{1})\big]
 \varepsilon^{\nu\,\ast}(k)\,,
\label{loop_amp}
\end{eqnarray}
\noindent where $k_{\mu}$, $ p_{1\mu}$, and $ p_{2\mu}$ are photon,
lepton, and anti-lepton momenta, respectively. The coefficients
$a_{1}$, $a_{2}$, $b_{1}$, and $b_{2}$ can be found in
Ref.~\cite{Kachanovich:2020xyg}, where they are presented in a way
suited for numerical implementation. 

In the case of the electron, the tree-level contribution can be
completely neglected. For the muon, the tree-level contribution is large
enough to be relevant, with the amplitude reading
\begin{eqnarray}
  \mathcal{A}_{\rm tree} = -\frac{e^{2} m_{\ell} \varepsilon_{\nu}^\ast
  (k)}{2 m_{W} 
  \sin\theta_{W}}\bigg[\frac{\overline{u}(p_{1})(\gamma^\nu \slashed{k} 
  +2\,p_{1}^\nu)v(p_2)}{t-m_\ell^2}-\frac{\overline{u}(p_1)(\slashed{k}
  \gamma^\nu 
  + 2\,p_2^\nu) v(p_2)}{u-m_\ell^2}\bigg]\, , \label{tree}
\end{eqnarray}
\noindent where squared invariant masses are denoted by the Mandelstam
variables $s=(p_{1} + p_{2})^{2}$, $t=(p_{1} + k)^{2}$, and
$u=(p_{2} + k)^{2}$.
To regulate the Z-boson pole, we used a  
Breit-Wigner propagator in the diagram classes (c-f) and (h)
shown in Fig.~\ref{Diagrams}. To this end our result with full
dependence  on the gauge parameters helped us to avoid spurious gauge
dependences related to the fact that the $h\to Z\gamma$ vertex is
gauge-dependent once the $Z$ is taken off-shell. 

%%%%%%%%%%%%%%%%%%%%%%%%%%%%%%%%%%%%%%%%%%%%%%%%%%%%%%%%%%%%%%%%%%
\newpage
\section{Results}

\begin{figure}[tb]
%\hrule
	\begin{center}
		\subfigure[t][]{\includegraphics[width=0.488\textwidth]{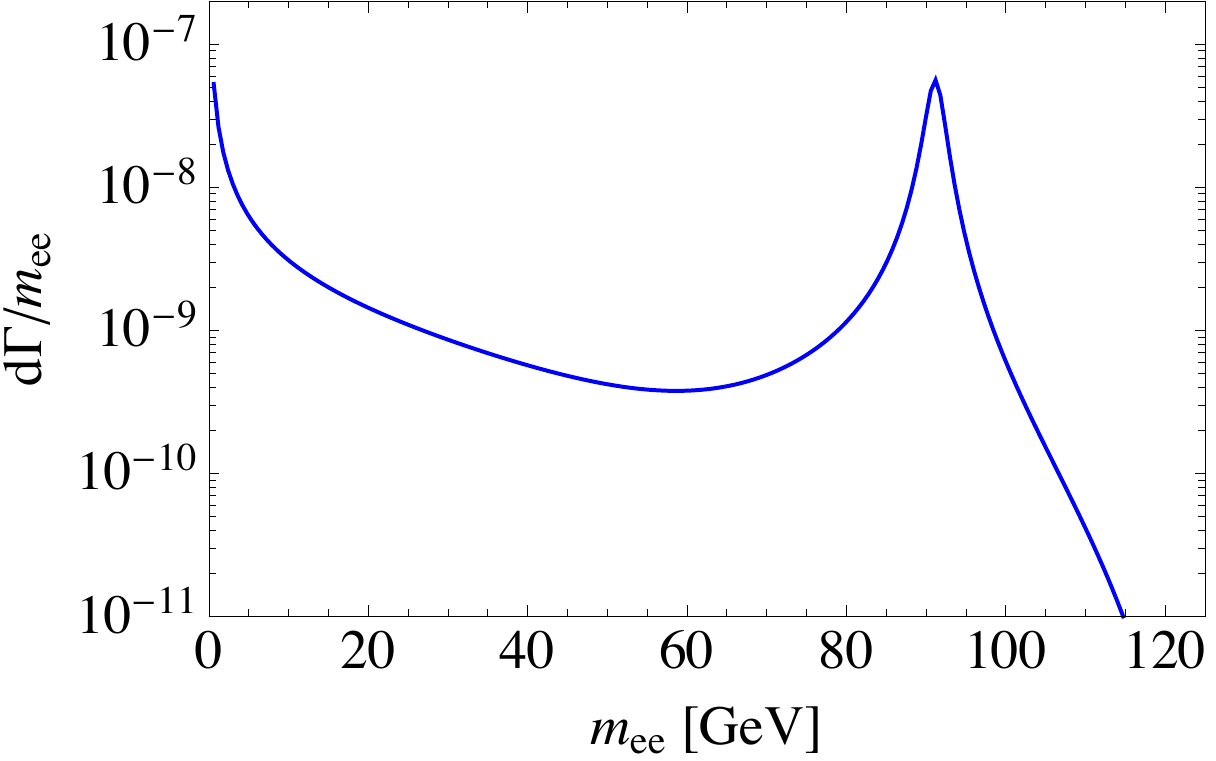}}
		\hspace{.1cm}
		\subfigure[t][]{\includegraphics[width=0.488\textwidth]{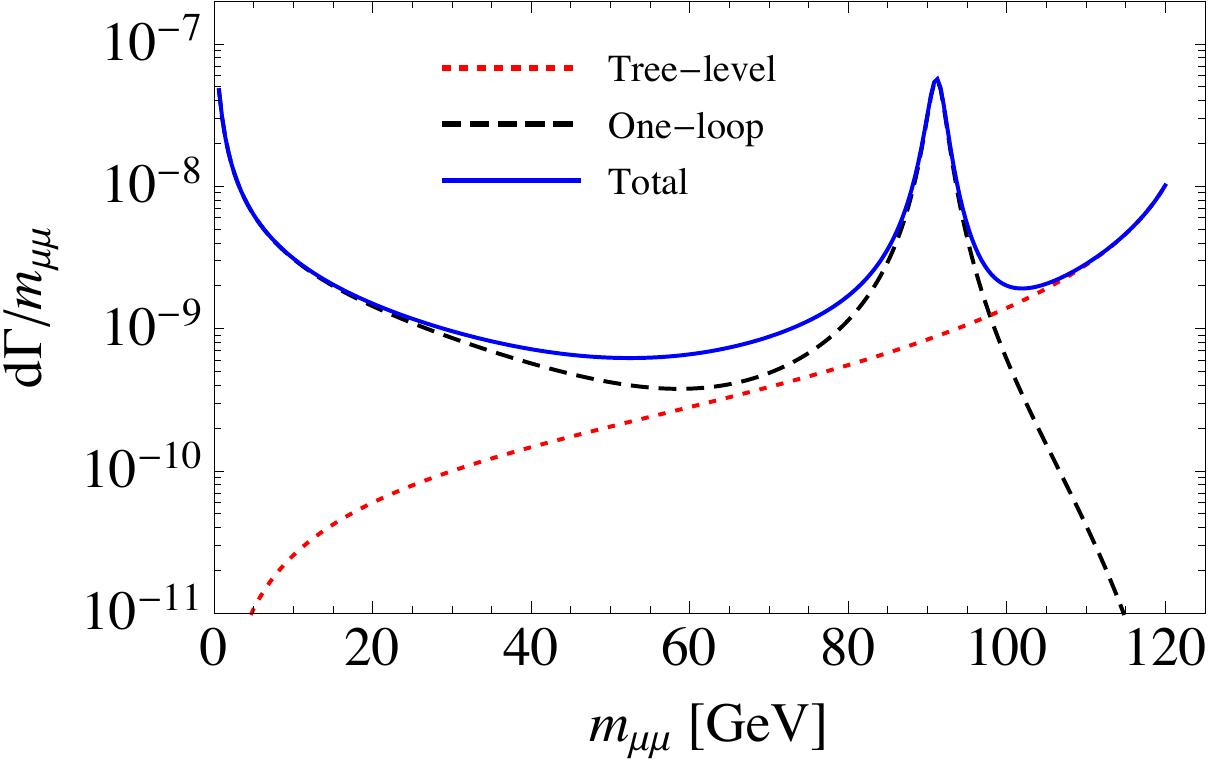}}
		%\hspace{.2cm}
		%\subfigure[t][]{\includegraphics[width=0.488\textwidth]{distribtaus.pdf}}
         \end{center}
         \caption{Differential decay rate for (a) electrons and (b)
           muons with respect to the invariant dilepton mass. The total,
           one-loop, and tree-level contributions are denoted by solid
           blue, black dashed, and red dotted lines, respectively. For
           the case of electrons, the tree-level contribution is
           negligible. The cut $E_{\gamma\text{,\,min}}=\,5\,\text{GeV}$
           has been implemented.}
	\label{Fig:3}
\hrule
\end{figure}

Our results have been obtained using following values of input
parameters:
\begin{equation}\label{Eq.:input paramiters}
\begin{split}
  &\qquad  m_{W} = 80.379\,
  \text{GeV}\,, \qquad m_{Z} = 91.1876\,\text{GeV}\,, \qquad \sin^2\theta_{W} = 1 - \frac{m_{W}^{2}}{m_{Z}^{2}}=0.223013\,,\\
&  
  m_{t} = 173.1\,\text{GeV} \,,\qquad m_{H} = 125.1\,\text{GeV}\,,\qquad m_{e} = 5.110\times 10^{-4}\,\text{GeV}\,,\qquad m_{\mu} = 0.106\,\text{GeV}\,,  \\
 &  \quad\quad G_{F} = 1.1663787\times 10^{-5}\,\text{GeV}^{-2}\,,\qquad \alpha^{-1} =
  \frac{\pi}{\sqrt{2}G_{F} m_{W}^{2} \sin^{2}\theta_{W}} = 132.184\,,
 	\end{split}
\end{equation}
\begin{figure}[tb]
%\hrule
	\begin{center}
		\subfigure{\includegraphics[width=0.52\textwidth]{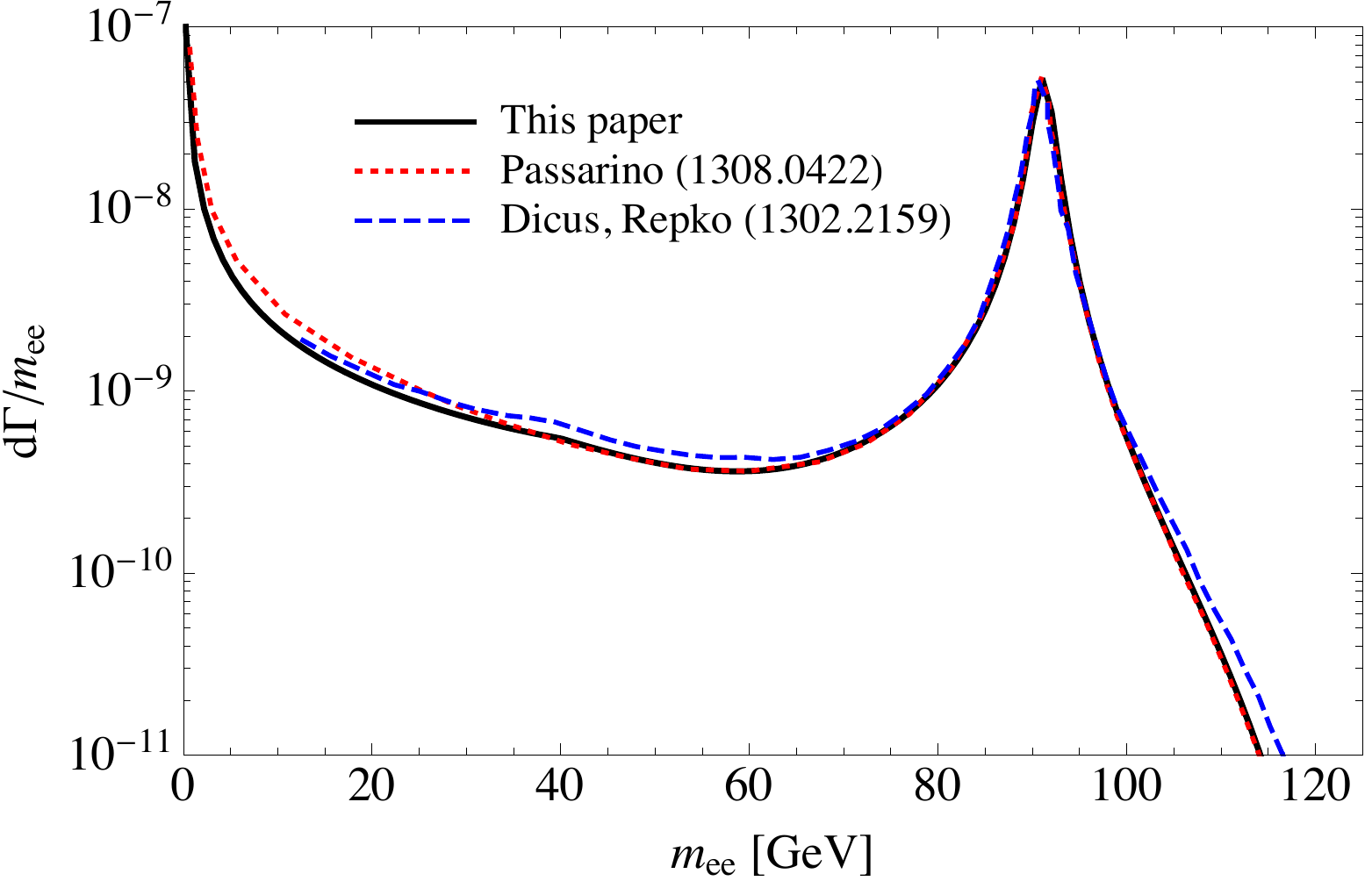}}
		\hspace{.1cm}
         \end{center}
         \caption{Differential decay rate with respect to the invariant
           dielectron mass. Our new result is denoted by a black solid
           line, while the results of Refs.~\cite{Dicus:2013ycd} and
           \cite{Passarino:2013nka} are denoted by blue long-dashed and
           red short-dashed lines, respectively.}\label{Fig:Comparison}
%[-3mm]\hrule
\end{figure}
\noindent % where $G_{F}$ is Fermi coupling constant.
The choice of the value for the inverse fine structure constant is
dictated by Eq.~\ref{Eq.:fine structure constant} and crucial for the
vanishing of the gauge dependence. Radiative corrections lead to the
more familiar value $\alpha^{-1} \simeq 128$, but the proper use of this
value would require the inclusion of higher-order corrections to the
decay amplitude as well.

The full decay rate has been evaluated after employing the kinematics
cuts of Ref.~\cite{Passarino:2013nka, Dicus:2013ycd}, namely:
\begin{align}
& s, t, u > (0.1\,m_H)^2,\qquad E_\gamma > 5\,\text{GeV},\nonumber\\
&  (E_{1} > 7 \,\text{GeV}\,,\qquad 
  E_{2} > 25\,\text{GeV})\quad\text{or}\quad (E_{1} > 25 \,\text{GeV}\,,\quad 
  E_{2} > 7\,\text{GeV})\,.  \nonumber
\end{align}
%\begin{figure}[tb]
%\hrule ~\\[2mm]
%	\begin{center}
%		\subfigure{\includegraphics[width=0.52\textwidth]{plots/Plovertmu.pdf}}
%         \end{center}
%         \caption{Differential decay rate with respect to the invariant mass
%           $\sqrt{t}\equiv m_{\mu\gamma}$ of the muon-photon pair.  The
%           tree-level, one-loop, and total contribution are denoted by
%           red dotted, black dashed lines, and solid blue lines,
%           respectively.  No cuts on $s$ are introduced.  }
%	\label{Fig:4}
%~\\[-3mm]\hrule
%\end{figure}

\begin{figure}[tb]
%\hrule
%~\\[2mm]
  \begin{center}
    \subfigure{\includegraphics[width=0.48\textwidth]{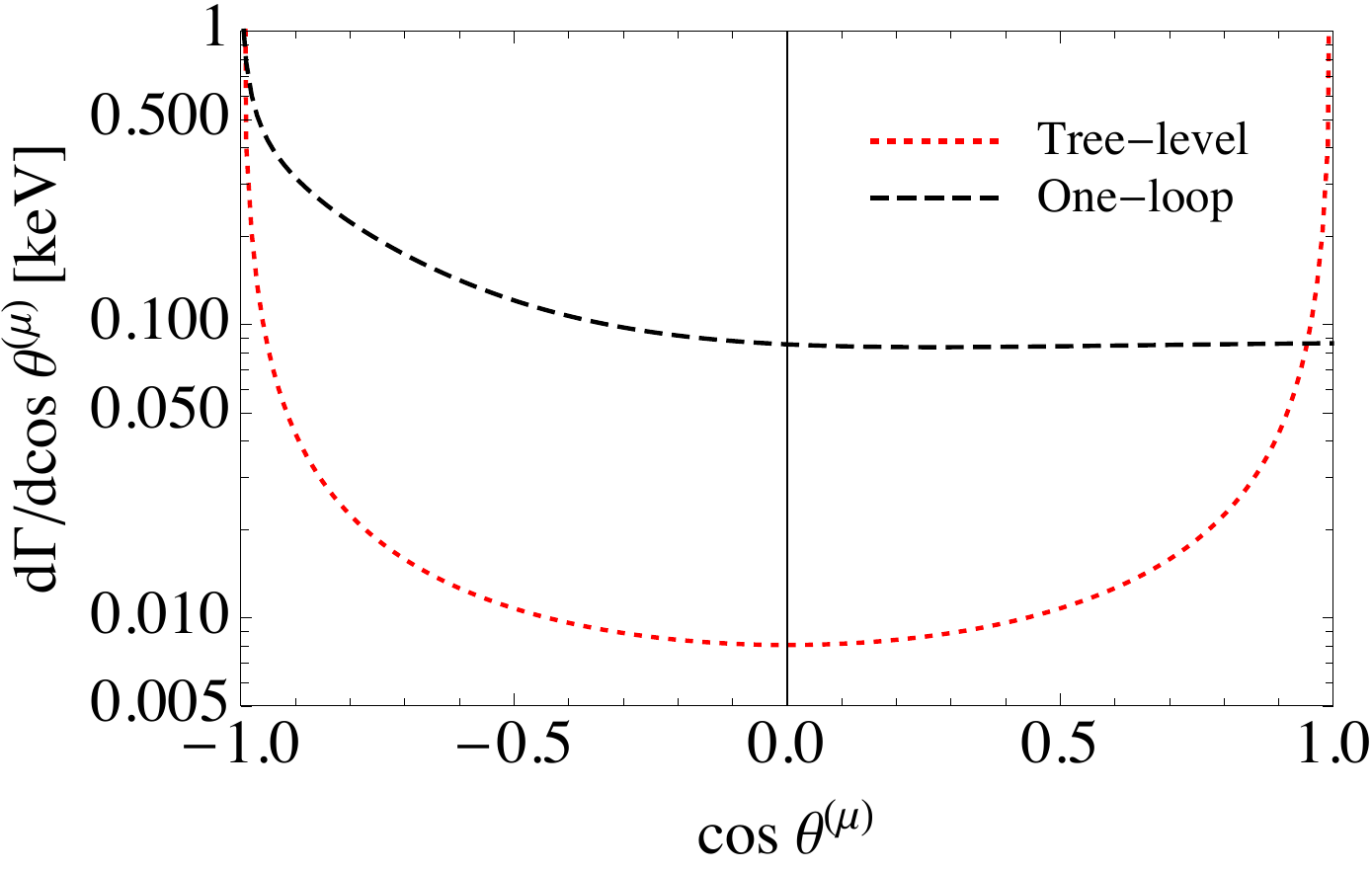}}
         \end{center}
         \caption{Differential decay rate with respect to
           $\cos\theta^{(\mu)}$, where $\theta^{(\mu)}$ is the angle
           between the photon and one of the leptons in the Higgs boson
           rest frame. The cuts $m_{\mu\mu} > 0.1 m_H$ and
           $E_{\gamma, \text{min}} = 5\,\text{GeV}$ have been
           implemented.  }
\label{Fig:FB-assymetry}
~\\[-3mm]\hrule
\end{figure}

We also predict the differential decay distribution with respect to the
angle between lepton and photon, $\theta^{(\ell)}$. The dependence of
the decay rate on $\cos \theta^{(\ell)}$ is shown in
Fig.~\ref{Fig:FB-assymetry}. The forward-backward asymmetry is defined
as
\begin{equation}\label{Eq.:higgstollgamma, Forward-backward assymetry}
\mathcal{A}_{l,\text{FB}}=\frac{\int_{-1}^0
  \frac{d\Gamma}{d\cos\theta_{l}} - \int_{0}^1 \frac{d\Gamma}{d\cos\theta_{l}}}{ 
  \int_{-1}^0 \frac{d\Gamma}{d\cos\theta_{l}} + 
  \int_{0}^1 \frac{d\Gamma}{d\cos\theta_{l}}}\,,
\end{equation}
\noindent where cuts $m_{\ell\ell} > 0.1 m_H$ and
$E_{\gamma, \text{min}} = 5\,\text{GeV}$ have been used.
We find:
\begin{equation}\label{Eq.:Forward-backward assymetry results}
  \mathcal{A}_{e,\text{FB}}=0.343\,,\hspace{10px}\qquad
  \mathcal{A}_{\mu,\text{FB}}=0.255\,.
\end{equation}

\section{Conclusions}
A novel analysis of the SM predictions for various observables in the
decays $h\to \ell \bar \ell \gamma$ has been presented. We have
clarified discrepancies in the literature and the issue of a
gauge-independent implementation of a finite $Z$ width.  Furthermore, a
gauge-independent result requires that the weak mixing angle defined in
terms of the gauge couplings fulfills the on-shell-scheme definition
$m_{W}=\cos \theta_{W} m_{Z}$.

For typical choices of cuts we find the branching ratios
$B(h\to e \bar e \gamma) = 5.8 \cdot 10^{-5}$ and
$B(h\to \mu \bar \mu \gamma) = 6.4\cdot 10^{-5}$ as well as the
forward-backward asymmetries $\mathcal{A}^{(e)}_{\text{FB}}=0.343$ and
$\mathcal{A}^{(\mu)}_{\text{FB}}=0.255$.  We provide analytic results in
Appendix A of Ref.\ \cite{Kachanovich:2020xyg} and in the corresponding
ancillary files.

\section*{Acknowledgements}
I thank Ulrich Nierste and Ivan Ni\v sand\v zi\'c for
the enjoyable collaboration on the presented work and
acknowledge the support from the doctoral
school \emph{KSETA}\ and the \emph{Graduate School Scholarship
  Programme}\ of the \emph{German Academic Exchange Service (DAAD)}.

\end{document}